\documentclass[prd,superscriptaddress,twocolumn,nopreprintnumbers,floatfix,nofootinbib]{revtex4}
\usepackage{url,amssymb,amsmath,longtable,rotating,units,wasysym,subfigure,epsfig,multirow,amsbsy,soul,mathrsfs,mathtools}
\usepackage[plainpages=false, colorlinks=true, anchorcolor=blue, linkcolor=blue, citecolor=blue, bookmarks=false]{hyperref}
\usepackage{float}
\usepackage{resizegather}
\usepackage{appendix}
\usepackage{bbold}
\usepackage{graphicx}
\usepackage{mathtools}
\usepackage[dvipsnames]{xcolor}
\usepackage{lipsum}
\usepackage{comment}
\usepackage{breqn}
\usepackage{autobreak}
\usepackage{empheq}

\usepackage[normalem]{ulem}

\newcommand{\SPA}{School of Physics and Astronomy, Monash University, VIC 3800, Australia}
\newcommand{\OzGravMonash}{OzGrav: The ARC Centre of Excellence for Gravitational Wave Discovery, Clayton VIC 3800, Australia}
\newcommand{\PennState}{Institute for Gravitation and the Cosmos, Department of Physics, Pennsylvania State University, University Park, PA, 16802, USA}
\newcommand{\MIT}{LIGO Laboratory, Massachusetts Institute of Technology, Cambridge, MA 02139, USA}
\newcommand{\Kavli}{Department of Physics and Kavli Institute for Astrophysics and Space Research, Massachusetts Institute of Technology, \\ 77 Massachusetts Ave, Cambridge, MA 02139, USA}
\newcommand{\Cardiff}{School of Physics and Astronomy, Cardiff University, Cardiff CF24 3AA, UK}
\newcommand{\PennStatePhys}{Department of Astronomy and Astrophysics, Penn State University, University Park PA 16802, USA}
\newcommand{\UWM}{Department of Physics, University of Wisconsin-Milwaukee, Milwaukee, WI 53201, USA}
\newcommand{\UMassDMath}{Department of Mathematics and
			Center for Scientific Computing and Visualization Research,
			University of Massachusetts, Dartmouth, MA 02747, USA}
			
\newcommand{\OzgGravANU}{OzGrav, ANU Centre for Gravitational Astrophysics, Research Schools of Physics, and Astronomy and Astrophysics, The Australian
National University, ACT, 2601, Australia}
\newcommand{\OzGravAdelaide}{OzGrav, University of Adelaide, Adelaide, SA 5005, Australia}

\begin{document}

\title{Bayesian inference for gravitational waves from binary neutron star mergers in third-generation observatories}

\author{Rory Smith}\email{rory.smith@monash.edu}\affiliation{\SPA}\affiliation{\OzGravMonash}
\author{Ssohrab Borhanian}\email{sub284@psu.edu}\affiliation{\PennState}
\author{Bangalore Sathyaprakash}\email{bss25@psu.edu}\affiliation{\PennState}\affiliation{\PennStatePhys}\affiliation{\Cardiff}
\author{Francisco Hernandez Vivanco}\affiliation{\SPA}\affiliation{\OzGravMonash}
\author{Scott E. Field}\affiliation{\UMassDMath}
\author{Paul Lasky}\affiliation{\SPA}\affiliation{\OzGravMonash}
\author{Ilya Mandel}\affiliation{\SPA}\affiliation{\OzGravMonash}
\author{Soichiro Morisaki}\affiliation{\UWM}
\author{David Ottaway}\affiliation{\OzGravAdelaide}
\author{Bram Slagmolen}\affiliation{\OzgGravANU}
\author{Eric Thrane}\affiliation{\SPA}\affiliation{\OzGravMonash}
\author{Daniel T\"{o}yr\"{a}}\affiliation{\OzgGravANU}
\author{Salvatore Vitale}\affiliation{\MIT}\affiliation{\Kavli}

\begin{abstract}
    Third-generation (3G) gravitational-wave detectors will observe thousands of coalescing neutron star binaries with unprecedented fidelity. Extracting the highest precision science from these signals is expected to be challenging owing to both high signal-to-noise ratios and long-duration signals. 
    We demonstrate that current Bayesian inference paradigms can be extended to the analysis of binary neutron star signals without breaking the computational bank. We construct reduced order models for $\sim\unit[90]{minute}$ long gravitational-wave signals, covering the observing band ($\unit[5-2048]{Hz}$),  speeding up inference by a factor of $\sim 1.3\times 10^4$ compared to the calculation times without reduced order models. The reduced order models incorporate key physics including the effects of tidal deformability, amplitude modulation due to the Earth's rotation, and spin-induced orbital precession.  We show how reduced order modeling can accelerate inference on data containing multiple, overlapping gravitational-wave signals, and determine the speedup as a function of the number of overlapping signals.
    Thus, we conclude that Bayesian inference is computationally tractable for the long-lived, overlapping, high signal-to-noise-ratio events present in 3G observatories.
\end{abstract}

\maketitle

\textit{{Introduction.---}} Third-generation (3G) gravitational-wave detectors such as Cosmic Explorer (CE) \cite{reitze2019cosmic} and the Einstein Telescope (ET) \cite{Maggiore2020} will observe hundreds of thousands to millions of binary neutron star (BNS) mergers a year \cite{Baibhav_2019, Samajdar}. Many of the observed signals will be extremely loud, with signal-to-noise ratios (SNRs) $\sim{\cal O}(100-1000)$. These signals will provide exquisite measurements of neutron star masses, tidal deformability, and spins, facilitating breakthroughs in cosmology and fundamental physics \cite{Vitale_2017, Adhikari_2019}. Analyzing signals in the 3G era will require scaling data analysis methods by orders-of-magnitude beyond their current capabilities: signals will be in band up to around 40 times longer than in Advanced LIGO/Virgo, the event rate will be thousands of times higher, and multiple signals will be in band at any one time \cite{reitze2019cosmic, Maggiore2020}.

Bayesian inference is the gold standard for measuring the properties of gravitational-wave signals \cite{Veitch_2015, Ashton:2018jfp, Vallisneri_2008, Smith_2020}. In Bayesian inference, the posterior probability density of source parameters $\Theta$ given experimental data $d$, and an hypothesis for the data $\mathcal{H}$, is:
\begin{equation}
p(\Theta|d, \mathcal{H}) = \frac{\pi(\Theta|\mathcal{H})\mathcal{L}(d|\Theta, \mathcal{H})}{\mathcal{Z}(d|\mathcal{H})}\,,
\end{equation}
where $\pi(\Theta|\mathcal{H})$ is the prior distribution, $\mathcal{L}(d|\Theta, \mathcal{H})$ is the likelihood function and $\mathcal{Z}(d|\mathcal{H})$ is the evidence. The posterior, $p(\Theta|d, \mathcal{H})$, is the target of parameter estimation, and the evidence is the target for hypothesis testing/model selection. As research and development of 3G instruments ramps up, there is increasing interest in the posterior density of gravitational-wave source properties because it is fundamental to answering interesting questions about the astrophysics capabilities of the detectors.
However, because of the high cost of computing the posterior density for BNSs, approximate methods are often used to study the capabilities of 3G detectors; see, e.g., \cite{Zhao_2018}. 
Fisher-matrix analyses have been used to approximate the width of $p(\Theta|d, \mathcal{H})$, assuming the likelihood is well approximated by a Gaussian distribution. While this assumption is valid for some projections of the posterior, it is not generally valid---even when the SNR is in the thousands---and must be carefully vetted~\cite{Vallisneri_2008}.  
Hence, a Bayesian treatment of parameter estimates is timely in order to reliably study topics in neutron star astrophysics with networks of 3G detectors.

In this \textit{Letter} we demonstrate how Bayesian inference can be performed on BNS signals using reduced order models (ROMs) \cite{Canizares:2014fya, PhysRevD.94.044031, Purrer:2014fza} of gravitational waveforms. Our work extends previous applications of reduced order modeling to signals that are up to 90 minutes in duration from a low frequency of $5\,$Hz, which is close to the expected low-frequency cutoff for 3G detectors \cite{reitze2019cosmic, Maggiore2020}. In addition, the ROMs incorporate effects of the Earth's rotation on GW signals, tidal deformability of neutron stars, and spin-induced orbital precession. We show that our ROMs are accurate representations of the original waveforms at around the $\sim10^{-12}$ level, ensuring that the ROM approximation is valid up to $\text{SNR}\approx 10^6$ \cite{PhysRevD.78.124020}---good enough for essentially all foreseeable 3G science.

The ROMs form highly efficient approximations of the likelihood function---the most expensive part of inference. This approximation is known as a \emph{reduced order quadrature} (ROQ) \cite{PhysRevD.94.044031, Canizares:2014fya, antil2013two}. We show that the ROQ can speed up the evaluation of the likelihood function by a factor of around 13,000 on individual BNS signals. This makes inference on these signals tractable. 
Additionally, we show that the ROM/ROQ framework can be applied to accelerate likelihood calculations on multiple overlapping in-band signals. At any given time, there are likely to be hundreds of BNSs in the Universe emitting gravitational waves in the 3G observing band~\cite{GW170817_stoch}. However, signals are sufficiently separated in time that they can usually be analysed separately, though, sometimes two or more signals are sufficiently close that a simultaneous analysis is required~\cite{Samajdar}. We determine the theoretical speedup factor for multiple in-band signals and show that the ROM/ROQ framework can speed up inference by a factor of around 10,000 for several overlapping signals.

We perform Bayesian inference on a $\unit[90]{min}$-long signal at a similar luminosity distance to GW170817, added into synthetic data of a 3G network consisting of CE, ET, and a southern-hemisphere CE-like detector which we call CE-South. The signal has SNR$\approx$2400. We are able to perform Bayesian inference in around 1600 CPU hours. Without the ROQ, the run time of the analysis would be around $10^7$ CPU hours. We overcome limitations of the Fisher information matrix by accurately determining the uncertainties of source parameters whose posterior densities are highly non-Gaussian.
Our results demonstrate that even loud 3G signals can be analysed with modest computational resources.
However, reduced-order methods are essential for controlling the computational cost.

\textit{The likelihood function and reduced order models.}---The most expensive part of evaluating the posterior probability is the likelihood function because it involves computing gravitational waveforms. The log likelihood function is \cite{Owen1999}
\begin{align}
\label{eq:likelihood}
    \ln\mathcal{L} &\propto
    -\frac{1}{2}\Big\langle d - h,\, d - h \Big\rangle \,\nonumber\\
     &=  \langle d, h\rangle - \frac{1}{2}\langle h, h \rangle
     -\frac{1}{2}\mathcal{Z}_n\,,
\end{align}
where the constant $\mathcal{Z}_n = \langle d, d \rangle$ is known as the ``noise evidence,'' and quantifies the likelihood of the data under the hypothesis that it is Gaussian noise \cite{Abbott_2020_LRR}. The angle brackets $\langle a,b\rangle$ denote the usual noise-weighted inner product \cite{PhysRevD.81.062003}.

In the frequeny domain, the gravitational waveforms $h$ have the general form~\cite{Zhao_2018}
\begin{dmath}
\label{eq:full_waveform}
h(f;\Theta) =  \frac{1}{D_L}\Big[F_{+}(f;\xi)\,h_{+}(f;\theta,\iota,\phi_c,t_c)+ F_{\times}(f;\xi)\,h_{\times}(f;\theta,\iota,\phi_c,t_c)\Big]\,,
\end{dmath}
where $h_{(+,\times)}(f;\theta,\iota,\phi_c,t_c)$ are the individual gravitational wave polarizations and are a function only of the \textit{intrinsic parameters} $\theta$, orbital inclination $\iota$, and phase and time at coalescence $\phi_c$ and $t_c$. $D_L$ is the luminosity distance to the source. The quantities $F_{(+,\times)}$ are the detector response functions, which depend on the binary's right ascension $\alpha$, declination $\delta$ and polarization phase $\psi$, which define $\xi=(\alpha, \delta, \psi)$. The full set of parameters $\Theta$ which appears on the left-hand side is the combined set of intrinsic and extrinsic parameters, i.e., $\Theta=(\theta, \xi ,\iota,\phi_c,t_c, D_L)$. The detector response functions are also functions of time/frequency due to the Earth's rotation, which cannot be neglected for BNS signals starting from $5\,\mathrm{Hz}$ \cite{Zhao_2018}. Since the time evolution of the response functions is slow compared to that of the gravitational-wave signal, we can use the stationary phase approximation which allows us to separate $F_{(+,\times)}$ and $h_{(+,\times)}$---see \cite{Zhao_2018} for details, and the explicit form of the frequency-dependent $F_{(+,\times)}$.

In the frequency domain, the ROMs of $h$~(Eq.~\ref{eq:full_waveform}) have the following general form \cite{PhysRevD.94.044031}:
\begin{eqnarray}
    \label{eq:rom}
     h^{\text{ROM}}(f;\Theta) &=& \sum_{J=1}^{N} h(\mathcal{F}_J;\Theta)\,B_{J}(f)\,\,.
\end{eqnarray}
The quantities $B_{J}(f)$ are a basis set which span the space of the signal. The $h(\mathcal{F}_J;\Theta)$ which appear on the right-hand side of Eq.~\ref{eq:rom} are the unapproximated waveform (Eq.~\ref{eq:full_waveform}) evaluated at a frequency $\mathcal{F}_J$ from a reduced set of $N$ frequencies $\lbrace\mathcal{F}_J\rbrace_{J=1}^{N}$. Previous work on reduced order modeling for gravitational-wave parameter estimation also constructed ROMs for the waveform amplitude $h^{*}h$ which is used to approximate the $\langle h, h \rangle$ term in the likelihood (Eq.~\ref{eq:likelihood}) \cite{PhysRevD.94.044031}. We find this unnecessary, and this term can be computed at negligible cost which we discuss below. 

The ROM requires $M/N$ fewer waveform evaluations than the unapproximated expression for the waveform, where $M = T\times\lbrack f_{\max}-f_{\min}\rbrack$ and $T$ is the signal duration. This is given by the Nyquist theorem, assuming a fixed sampling rate. For 3G detectors, we assume $f_{\max}=\unit[2048]{Hz}$, $f_{\min}=\unit[5]{Hz}$, $T=\unit[90]{min}\,(\unit[5400]{s})$. When ROMs are substituted into the likelihood function, they form a compressed inner product known as a reduced order quadrature (ROQ). The ROQ speeds up the likelihood function by a factor \cite{Canizares:2014fya, PhysRevD.94.044031}
\begin{equation}
    S \approx M/N\,,
    \label{eq:speedup}
\end{equation}
and in general $S \gg 1$. This speedup assumes that the waveforms have a closed-form expression, which frequency-domain waveforms typically do.

\textit{ROM construction.}---The ROMs are constructed in three steps: $(i)$ make a representative ``training-space'' of gravitational waveforms which span the parameter range of interest; $(ii)$ select basis elements from the training set; and $(iii)$ determine the reduced set of frequency nodes. All steps are achieved using a greedy algorithm \cite{greedycpp, PhysRevX.4.031006, BARRAULT2004667}. We construct ROMs of $\unit[90]{min}$-long gravitational-wave signals including spin precession and tidal deformability. That we can build ROMs for $\unit[90]{min}$-long BNS signals should not be taken for granted. Previous studies \cite{Canizares:2014fya, PhysRevD.94.044031, Morisaki_2020} have not established whether ROMs for such signals can be made in practice or if they would be practical for data analysis. In \cite{Canizares:2014fya, PhysRevD.94.044031, Morisaki_2020} various scalings for ROM basis sizes are given as a function of low-frequency of the signals, or parameter space ranges. However, there has been no systematic study on the size of ROM bases on both low-frequency \textit{and} parameter space size. 
The fundamental issue is whether the parameter space can be made small enough to be both effective and efficient for long-duration signals. Below, we show that it is indeed the case. 

We focus on individual signals, and target a small region of intrinsic parameter space on which we build a training set. The parameters are the chirp mass $\mathcal{M}_c$, symmetric mass ratio $\eta$, tidal deformabilities $(\lambda_1, \lambda_2)$, spin components projected along the orbital angular momentum axis $(\chi_1^z, \chi_2^z)$, effective-precession spin and the initial value of the azimuthal precession angle $(\chi_p, \alpha_0)$, and orbital inclination $\iota$. We use the waveform model \textsc{IMRPhenomPv2$\_$NRTidalv2}~ \cite{PhysRevD.99.024029, PhysRevD.96.121501, PhysRevD.100.044003}, which is parameterized by the vector $\theta = \{\mathcal{M}, \eta, \chi^z_1, \chi^z_2, \chi_p, \alpha_0, \iota, \lambda_1, \lambda_2\}$. Inclination appears here because it evolves during the inspiral due to spin-induced orbital precession. 

The size of the ROM basis is sensitive to the range in chirp mass. We pick a fiducial chirp mass value of $\mathcal{M}_{*}=1.385\,M_{\odot}$ and restrict the width of the chirp mass of the training set to be $\pm 5\times10^{-4}M_{\odot}$. This mass range is approximately $1\times10^{3}\Delta\mathcal{M}$, where $\Delta\mathcal{M}$ is the Fisher-matrix error, estimated using \texttt{gwbench} \cite{Borhanian:2020ypi}. Following the Fisher-matrix error treatment in \cite{Poisson_1995}, we find that signals with SNRs of around 10 have Fisher errors $\Delta\mathcal{M}\sim10^{-4}M_{\odot}$.  Hence, our chirp mass range ensures that we can analyze signals with SNRs around 10 without artificially railing against prior bounds in mass. In practice, we may want to use broader priors in mass than are possible with a single parameter-space patch. Broader prior ranges can be employed simply by utilizing multiple ROM bases that individually span small parameter-space ranges. Around 1000 such patches in $\mathcal{M}-\eta$ would be needed to cover the full BNS mass space, assuming BNSs have chirp masses approximately in the range $1M_{\odot}$--$2M_{\odot}$.

All other intrinsic parameters are chosen to have physically-motivated ranges. The symmetric mass ratio is restricted to $0.2 \leq \eta \leq 0.25$. Assuming a minimum neutron star mass of $1 M_{\odot}$, this range ensures we describe neutron stars with masses up to $2.6 M_{\odot}$ (around the maximum plausible non-rotating neutron star mass \cite{190814, Essick_2020, 2019EPJA...55..209L}). For all other parameters, we consider the following ranges: $0 < \chi_{p} \leq 0.1$, $-0.1 \leq \chi_{1,2}^z \leq 0.1$, $0 \leq \lambda_{1,2} \leq 5000$. Additionally, the ROM is constructed to be valid for all values of sky location parameters $\xi=(\alpha, \delta, \psi)$, luminosity distance $D_L$, phase at coalescence $\phi_c$, $\iota$, and $\alpha$.  We consider three starting frequencies  $f_{\min} = \unit[5, 10, 20]{Hz}$, maximum frequency $f_{\max}=\unit[2048]{Hz}$, and signal duration of $T=\unit[90]{min}$. These values of $f_{\min}$ test how the size of the ROM bases scales with the low-frequency cut off.

We construct a training set of waveforms for the parameter space defined above. The basis and reduced frequency nodes are selected using a greedy algorithm. Details about the training set and greedy algorithm are described in the supplementary material. For signals starting in band from $\unit[5, 10, 20]{Hz}$, the ROMs have $N=522, 291, 179$ basis elements. The basis size only increases by a factor of three when going from $\unit[20]{Hz}$ to $\unit[5]{Hz}$, despite the signals being over 40 times longer in duration. Bases of around $\sim$ 500 elements should be typical for ROMs of BNS signals starting from $\unit[5]{Hz}$ with parameter ranges similar to those used here. Reducing the chirp mass to that of a $1 M_{\odot}+1 M_{\odot}$ binary will change the signal duration only by a factor of 2, much less than the difference in the duration of signals starting from $\unit[20]{Hz}$ vs $\unit[5]{Hz}$. Hence, the basis size should be roughly constant for lower-mass systems. 

The computational cost of building the ROM is relatively small. We require 160 16-core 2.20GHz Intel Xeon E5-2660 CPUs running for around 7 minutes, and then a single CPU running for around 2 hours to complete the basis construction -- see \textit{Step-(ii)} of the ROM building strategy in the Supplementary material. The memory footprint of the basis is around 90GB. Thus, it would be feasible to build reduced order models covering the full chirp-mass range of BNSs.

\textit{Likelihood speedup.}---The most efficient use of ROMs in Bayesian inference is to compress the large inner products in the likelihood function. The compressed inner products are known as a ROQ integration rule. We obtain the ROQ likelihood by substituting the ROM~(Eq.~\ref{eq:rom}) into the likelihood~(Eq.~\ref{eq:likelihood}). 
The ROQ likelihood is
\begin{equation}
    \ln\mathcal{L}_{\text{ROQ}} \propto L(\Theta) - \frac{1}{2}Q(\Theta) - \frac{1}{2}\mathcal{Z}_n\,,
    \label{eq:likliehood_roq}
\end{equation}
where the functions $L(\Theta)$ and $Q(\Theta)$ are given by
 \begin{align} 
 \label{eq:linear}
    L(\Theta) &=  \Re\sum_{J=1}^{N}h(F_J;\Theta)\,\omega_{J}(t_c)\,,\\
    Q(\Theta) &= \sum_{I=1}^{N}\sum_{J=1}^{N} h^{*}(F_I;\Theta)h(F_J;\Theta)\psi_{IJ}\,.\label{eq:quad}
\end{align}

\noindent The quantities $\omega_{J}(t_c)$ and $\psi_{IJ}$ are integration weights that depend only on the basis functions, data, and noise power spectral density, and are defined in the supplementary material. 

The computational cost of the ROQ likelihood scales as
\begin{equation}
\ln\mathcal{L}\sim\mathcal{O}(N \times W)\,,
\end{equation}
\noindent where $W$ is the number of operations required to evaluate the waveform at a given frequency.
Unlike previous work \cite{PhysRevD.94.044031}, we have chosen to write the $Q$ term without the use of an explicit basis for the waveform amplitude $h^{*}h$. The scaling of Eq.~\ref{eq:quad} is independent of $W$ because waveforms at the reduced frequencies have already been computed as part of $L$ (Eq.~\ref{eq:linear}). Thus, Eq.~\ref{eq:quad} scales like $\sim{O}(N^2)$ and we find that $N$ is small enough such that $N^2 \ll N\times W$. For our basis starting from $\unit[5]{Hz}$ (which contains $N=522$ basis elements), the theoretical speedup (Eq.~\ref{eq:speedup}) is $S=5400\text{s}\times(2048-5)\text{Hz}/522 \approx 21,000$. Empirically, we find a speed up of around 13,000. The degradation in performance is due to fixed overheads, such as allocating data structures for the waveforms. The integration weights $\omega_{IJ}(t_c)$ and $\psi_{IJ}$ are dependent on the data and noise power spectral density and have to be computed before data analysis can take place. 
The cost of computing both is negligible in practice. 

\textit{Validation and accuracy.}---The accuracy of the ROQ likelihood (Eq.~\ref{eq:likliehood_roq}) is limited by the accuracy of the ROM. We validate the accuracy by computing the mismatch $M$ between the ROM representation of $h$ (Eq.~\ref{eq:rom}) and its unapproximated form (Eq.~\ref{eq:full_waveform}): 
\begin{equation}
    M(h) = 1 - \frac{\langle h^{\text{ROM}}, h \rangle}{\sqrt{\langle h^{\text{ROM}}, h^{\text{ROM}} \rangle \langle h, h \rangle}} .
\end{equation}

\begin{figure}[ht]
  \centering
 \includegraphics[width=0.9\columnwidth]{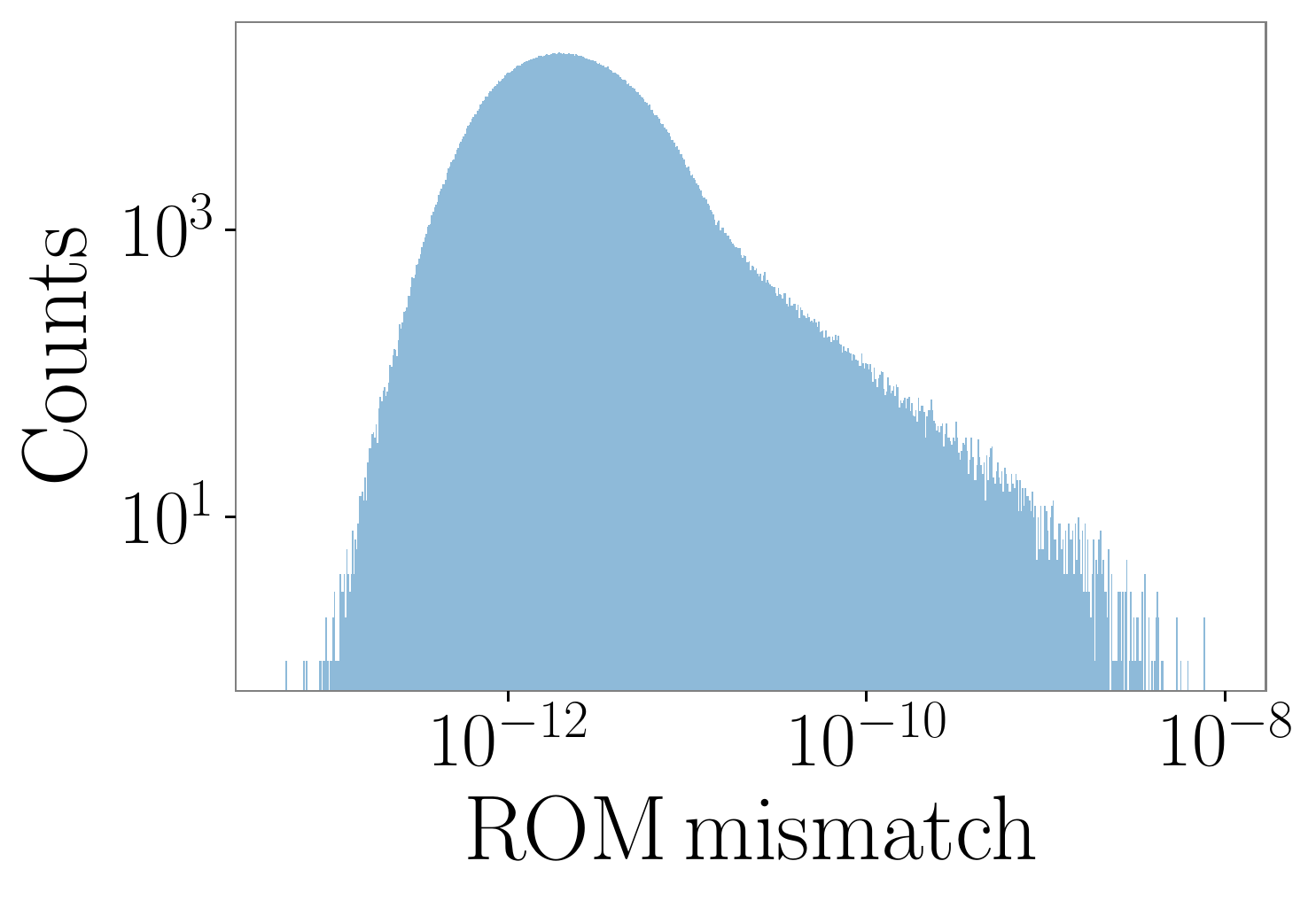}\quad
 \caption{Accuracy of the ROM approximation for 90-minute long BNS signals starting from $\unit[5]{Hz}$. The signals include amplitude modulation due to the Earth's rotation. }
 \label{fig:mismatch}
\end{figure}

\noindent In the noise-weighted inner products, we assume a flat power spectrum, meaning our mismatches are more conservative than if one used a gravitational-wave detector noise power spectral density. In Fig.~\ref{fig:mismatch} we show the mismatch $M(h)$ for $2\times10^6$ random parameter values $\Theta$ that were not included in the training space. We include random sky locations, inclinations, luminosity distances, and phases at coalescence. We also include the frequency-dependent response functions in the mismatch calculations, demonstrating that the ROM is accurate describing signals with amplitudes modulated by the Earth's rotation. The mismatch is strongly peaked around $10^{-12}$, ensuring that parameter estimation will be unbiased up to $\text{SNR}\approx 10^6$ (so that twice the mismatch multiplied by the SNR squared is less than unity \cite{PhysRevD.78.124020, Ohme_2012}).

\textit{Inference with a high-SNR signal.}---As an illustrative example, we consider inference on a signal, which we nickname GW370817.
The parameters are: $\Theta_{370817}=\{\mathcal{M}=1.3854\,M_{\odot}, \eta=0.24925, \chi^z_{1}=-0.0113, \chi^z_{2}=0.01070, \chi_p=0.03, \alpha_0=1.1, \iota=0.785, \lambda_1=422.5, \lambda_2=839.4, D_L=\unit[38.77]{Mpc}, \alpha=\unit[1]{hours}\, \unit[57]{min}\, \unit[20.5]{s}, \delta=\unit[-14.9]{deg}, \psi=2.012, \phi_c=0\}$. This signal has a luminosity distance consistent with GW170817. We add the signal into synthetic data of a three-detector network consisting of CE, ET, and CE-South. We use a ``zero-noise'' realization of Gaussian noise \cite{Pankow_2018}, which (statistically) is the most likely realization. This noise realization has the added convenience that if we use flat priors, the posterior peaks at exactly the true parameter values which serves as a useful diagnostic check. The signal has $\text{SNR}=2400$. We use flat priors on all parameters, and the ranges are given by the range of validity of the ROM. In addition, we impose a physically motivated prior constraint on the component tidal deformability: $\lambda_2 > \lambda_1$. Lastly, we use a uniform prior over a 0.2s interval centered on the true trigger time. In general it is not necessary to restrict the chirp mass prior to such a narrow range. Provided that ROMs exist in local patches covering an extended chirp-mass region, a wide prior can be utilized by building ROQ weights from multiple ROM bases.

\begin{figure}
\centering
\includegraphics[width=1\columnwidth]{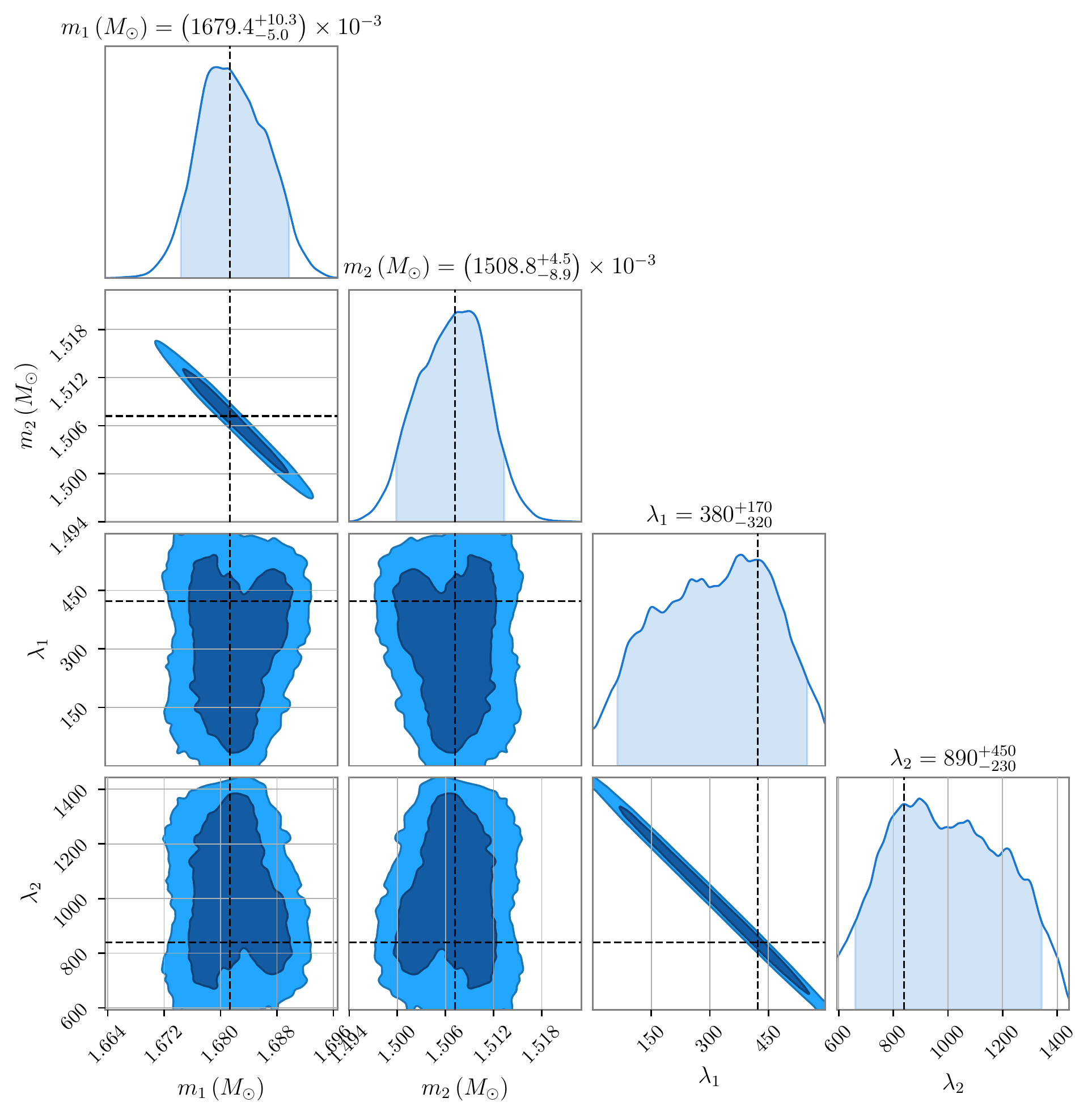}
\caption{One- and two-dimensional posterior densities for component masses and component tidal deformabilities. Dark/light shading indicates the one-sigma/two-sigma credible interval respectively. True parameter values are indicated by dashed lines.
}
\label{fig:corner1}
\end{figure}

We use the \texttt{dynesty} nested sampling package to infer the posterior density. In order to obtain well-converged posteriors, we set the number of live points to 5000, and use a random-walk proposal from the \texttt{bilby} \cite{bilby, Smith_2020} inference library, which takes a number of steps equal to 70 times the running estimate of the autocorrelation length. The analysis is parallelized over 160 cores. The analysis takes $\unit[10]{hours}$ (1600 CPU hours) on a cluster of 16-core 2.20GHz Intel Xeon E5-2660 CPUs. The large CPU time occurs because the implementation of the nested sampling algorithm in \cite{bilby, Smith_2020} is extremely slow to converge when the SNR is in the thousands. However, only a handful of events are likely to be detected at these SNRs, with the vast majority of signals having ``moderate'' SNRs less than 100. Analysis on signals with moderate SNRs takes on the order of a day using a single CPU using ROM/ROQ techniques \cite{PhysRevD.94.044031}. In contrast, the CPU time without ROM/ROQ methods would be on the order of $\unit[20\times10^6]{hours}$, i.e., prohibitively expensive. This analysis highlights the need for improvements to the convergence of stochastic-sampling based approaches to inference. 

In Fig.~\ref{fig:corner1}, we show the one- and two-dimensional posterior densities for the component masses and tidal deformabilities.
The component masses can be constrained to the $\sim 5\times 10^{-3}M_{\odot}$ level at the 90$\%$ credible intervals, which is consistent with Fisher information estimates. The tidal deformabilities have broad uncertainty and are highly non-Gaussian, demonstrating the importance of full Bayesian inference for understanding how well tidal effects---and hence nuclear physics---can be constrained by 3G observatories.%

\textit{Overlapping signals.}---When multiple signals are present simultaneously, the log likelihood function is
\begin{align}
\label{eq:likelihood_mult}
    \ln\mathcal{L} &\propto
    -\frac{1}{2}\Big\langle (d - \sum_{i=1}^{n} h_i),\, (d - \sum_{j=1}^{n} h_j) \Big\rangle \,\nonumber\\
     &=  \Big\langle d, \sum_{i=1}^{n} h_i\Big\rangle - \frac{1}{2}\sum_{ij}\langle h_i, h_j \rangle
     -\frac{1}{2}\mathcal{Z}_n ,
\end{align}
where the sums run over the independent gravitational-wave signals. The double sum $\sum_{ij}$ runs over all pairs $(i,j)$. In the multiple-signal case, the ROQ likelihood and its scaling are 
\begin{align}
\label{eq:roq_mult}
    \ln\mathcal{L}_{\text{ROQ}} &= \sum_{i=1}^{n}L(\Theta_i) - \sum_{i=1}^{n}\frac{1}{2}Q(\theta_i)\nonumber\\ &+\sum_{\{j,k\}}^{n^2/2}R_{jk}(\Theta_j,\Theta_k, \Delta t^{jk}) - \frac{1}{2}\mathcal{Z}_n\,\\
    &\sim\mathcal{O}\Big(N\times n\times W + N^2\times(n+n^2/2)\Big)\,,\nonumber
\end{align}
where $L$ and $Q$ are given by Eq.~\ref{eq:linear} and Eq.~\ref{eq:quad}. The $R_{jk}$ term sums over products of all pairs of waveforms $\{j,k\}$, with $j>k$, and is also a function of the relative time offset between two signals, $\Delta t^{jk}$. The function $R_{jk}(\Theta_j,\Theta_k,\Delta t^{jk})$ is 

\begin{align}
    &\Big\langle h(\Theta_j), h(\Theta_k)\Big\rangle \approx R_{jk}(\Theta_j,\Theta_k, \Delta t^{jk}) =\nonumber\\ &\Re\sum_{K=1}^{N}\sum_{L=1}^{N} h^{*}(F_K;\Theta_j)h(F_L;\Theta_k)\Gamma_{KL}(\Delta t^{jk})\,,
    \label{eq:R}
\end{align}
where the matrix $\Gamma_{KL}(\Delta t^{jk})$ is a set of integration weights, given in the supplementary material. The overall speedup of the multiple-signal ROQ likelihood Eq.~\ref{eq:roq_mult} with respect to non-ROQ likelihood (Eq.~\ref{eq:likelihood_mult}) is
\begin{equation}
  S \approx \frac{MnW + Mn^2/2}{NnW +Nn^2/2}\,,\,\, \text{with $n>1$} \,,
  \label{eq:speedup_multi}
\end{equation}
\noindent where we have kept terms at $\mathcal{O}(W)$ and $\mathcal{O}(n^2)$.
The $N^2n^2$ scaling is potentially problematic if the number of in-band signals is large. However, most overlapping signals---roughly between $96\%$ to $99.5\%$---are well separated in time so that they can be analysed separately (tables I and IV of \cite{Samajdar}). Thus, we only consider the speedup for the simultaneous analysis of a few in-band signals. We empirically determine a speed-up of around $\sim\mathcal{O}(10^4)$ compared to the calculation time without ROM for up to 10 in-band signals. We show the speedup as a function of the number of in-band signals in the supplementary material.

\textit{Discussion.}---Reduced order models of gravitational-wave signals from BNS mergers can be used to accelerate parameter estimation in 3G observatories, thereby removing a computational hurdle. This work lays the groundwork for detailed studies of BNS systems in the 3G era. Further, the ROM/ROQ framework can be used to efficiently carry out inference on data containing multiple overlapping signals. Further avenues to pursue include ROMs of more sophisticated waveforms, e.g., with higher-order gravitational-wave modes, which can place tighter constraints on parameter estimates \cite{bustillo2020higherorder}. Bayesian inference on very loud signals---SNR $\sim\mathcal{O}(1000)$---requires significantly more likelihood evaluations than in analyses on LIGO/Virgo signals. Sampling-based methods for Bayesian inference will have to be significantly adapted and scaled up in order to efficiently analyze data in the 3G era. For instance, Hamiltonian Monte Carlo methods \cite{betancourt2018conceptual} (which exploit gradients of posterior densities) and machine learning techniques, e.g.,
\cite{gabbard2020bayesian, shen2019deterministic, PhysRevLett.124.041102, PhysRevLett.122.211101} (which provide rapid approximations to reduced order models and posterior densities) may be promising avenues to explore.

\textit{{Acknowledgements.}}---
This work was
supported through Australian Research Council (ARC) Centre of Excellence CE170100004. PL is a recipient of the Australian Research Council Future Fellowship
FT160100112 and Discovery Project DP180103155.
IM is a recipient of the Australian Research Council Future Fellowship FT190100574. ET is a recipient of the Australian Research Council Future Fellowship FT150100281. S.V. acknowledges support of the National Science Foundation through the NSF award PHY-1836814. We greatly appreciate suggestions and comments from members of the LIGO/Virgo Collaboration and Cosmic Explorer Consortium. In particular, Leo Tsukada, Nathan Johnson-McDaniel, Philippe Landry, and Aidan Brooks. This manuscript has LIGO Document number P2100051.

\section{Supplementary Material}

\subsection{ROQ integration weights}
The term $\omega_J(t_c)$ in Eq.~\ref{eq:linear} is a data-dependent integration weight, which depends only on the basis functions, data, noise power spectral density and coalescence time $t_c$ \cite{PhysRevD.94.044031} 
\begin{equation}
    \omega_J(t_c) = 4\Delta f\sum_{l=1}^{M} \frac{d^{*}(f_l)B_J(f_l)e^{-2\pi i f_l t_c}} {S_n(f_l)}\,.
    \label{eq:omega}
\end{equation}
\noindent Note that the above expression is similar to the noise-weighted inner product between the data and time-shifted basis, but the prodct differs in that we do not take only the real part. These data-dependent weights are a one-time, upfront calculation and can be efficiently computed using an inverse fast Fourier transform.

Similarly, the term $\psi_{IJ}$ in (Eq.~\ref{eq:quad}) is a data-dependent integration weight, which only depends on the basis and noise power spectral density \cite{PhysRevD.94.044031}:

\begin{equation}
    \psi_{IJ} = 4\Delta f\sum_{l=1}^{M}\frac{\,B^{*}_{I}(f_l)B_{J}(f_l)}{S_n(f_l)}\,.
    \label{eq:Psi}
\end{equation}

The term $\Gamma_{KL}(\Delta t^{jk})$ which appears in Eq.~\ref{eq:roq_mult} and Eq.~\ref{eq:R} is also a data-dependent integration weight, and depends on the basis, noise power spectral density and the relative time offset between two gravitational-wave signals $\Delta t^{jk}$. This weight appears in the ``cross-term'' in the likelihood function and is related to the correlation between two overlapping signals. It is given by
\begin{equation}
   \Gamma_{KL}(\Delta t^{jk}) = 4\Delta f \sum_{l=1}^{M} \frac{B^{*}_{K}(f_l)B_{L}(f_l)e^{-2\pi\,if_l\Delta t^{jk}}}{S_n(f_l)}\,.
    \label{eq:Gamma}
\end{equation}

\subsection{Summary of ROM building strategy}

The ROMs (Eq.~\ref{eq:rom}) are constructed from a ``training set'' of waveforms (Eq.~\ref{eq:full_waveform}) distributed on the signal's 9-dimensional $\textit{intrinsic}$ parameter space $\theta$. The ROMs are constructed in three steps: 

\begin{itemize}
    \item Step-$(i)$ Construct a training set of gravitational waveforms. 
    \item Step-$(ii)$ find the basis $\{B_i\}_{i=1}^{N}$ (\ref{eq:rom}) using a greedy algorithm \cite{greedycpp}.
 \item Step-$(iii)$ find the frequency nodes associated with the ROM $\{\mathcal{F}_i^l\}_{i=1}^{N}$ using the empirical interpolation method \cite{BARRAULT2004667, PhysRevX.4.031006}. 
\end{itemize}

Unlike previous work which considered ROMs for aLIGO/Virgo networks, we in principle also have to consider the effects of the Earth's rotation via the frequency-dependent detector response functions---see Eq.~\ref{eq:full_waveform}. Including these response functions means that the training space needs to cover the full set of sky locations and polarization angles which appear in Eq.~\ref{eq:full_waveform}. However, in practice we find that we can simply ignore the response functions' frequency-dependence in the training set for the ROM. Despite this omission in the construction, the ROM is nevertheless accurate at describing the full signal (Eq.~\ref{eq:full_waveform}) which includes the effects of frequency-dependent detector response functions. This is convenient, but also unsurprising: the effect of the amplitude modulation due to the Earth's rotation is a small perturbation on the gravitational-wave driven evolution of the signal's amplitude amplitude.  

For \textit{Step}$(i)$, we find that in practice we only need to build a training set consisting of $h_+$ and $h_{\times}$. We use the following strategy to construct effective training sets that sample the boundary and bulk of the parameter space. We split training space parameter samples into three categories: $(i)$ $\theta^{\text{boundary}}$ which are drawn from random values on the boundary of the space, $(ii)$ $\theta^{\text{grid}}$ which are drawn from random values on a uniform grid of 500 points in each parameter dimension, and $(iii)$ $\theta^{\text{bulk}}$ which are drawn from a uniform random distribution between the minimum and maximum parameter values. We find that equal numbers of each of $(\theta^{\text{boundary}}, \theta^{\text{grid}}, \theta^{\text{bulk}})$ works well  in practice. The training set is then simply an array 
\begin{eqnarray*}
T = \Big\{&h_+(f;\theta_1)&, h_{\times}(f;\theta_1),\,\\
&h_+(f;\theta_2)&, h_{\times}(f;\theta_2),\,,
\ldots\,,\\ 
&h_+(f;\theta_n)&, h_{\times}(f;\theta_n) \Big\}\,,
\end{eqnarray*}

\noindent where n is half the size of the training space. We use 4.3 million waveforms in the training set, which we find gives us an accurate basis without having to perform any additional refinement \cite{PhysRevD.94.044031}. We adaptiveley down-sample the waveforms in the training set as in \cite{PhysRevD.94.044031}, and upsample the basis to a fixed $\Delta f$ in postprocessing. In \textit{Step}$(ii)$, the elements of the basis set $\{B_i\}_{i=1}^{N}$ are simply judiciously chosen waveforms from the training set which are picked using a greedy algorithm as in \cite{greedycpp}. This processes is highly parallelizable. Once the basis has been found, it is then ``up-sampled'': re-computed at a fixed frequency resolution $\Delta f$. Up-sampling is a serial operation because it involves orthonormalizing the basis using the iterative and modified Gram-Schmidt algorithm \cite{greedycpp}. The frequency nodes (\textit{Step}$(iii)$) are found using the empirical interpolation method, which is also described and implemented in \cite{greedycpp}.

\subsection{The ROQ speedup for  multiple signals}
\begin{figure}
\centering
\includegraphics[width=1\columnwidth]{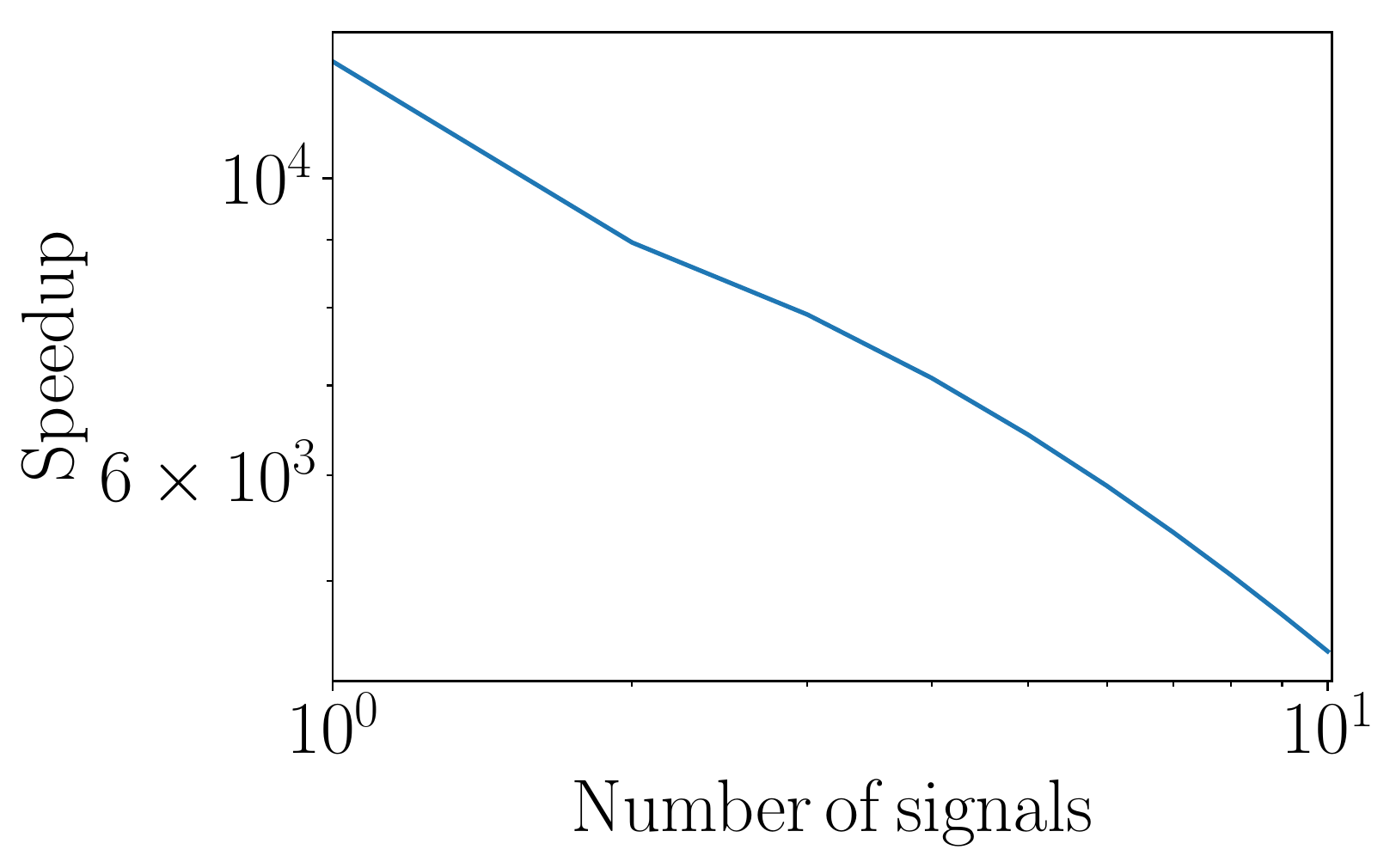}
\caption{ROQ-likelihood speedup (Eq.~\ref{eq:roq_mult}) for signals starting from 5Hz as a function of the number of overlapping in-band signals. The ROQ uses $N=522$ basis elements.}
\label{fig:speedup}
\end{figure}

In Fig.~\ref{fig:speedup} we show the empirically computed speedup (Eq.~\ref{eq:speedup_multi}) as a function of the number of in-band signals. We time the ROQ likelihood function (Eq.~\ref{eq:roq_mult}) and compare it to the non-ROQ version (Eq.~\ref{eq:likelihood_mult}) using one node of a single 2.20GHz Intel Xeon E5-2660 CPU.  The speedup is independent of the relative time offset between overlapping signals $\Delta t^{ij}$.

\bibliography{biblio}

\begin{thebibliography}{41}
\expandafter\ifx\csname natexlab\endcsname\relax\def\natexlab#1{#1}\fi
\expandafter\ifx\csname bibnamefont\endcsname\relax
  \def\bibnamefont#1{#1}\fi
\expandafter\ifx\csname bibfnamefont\endcsname\relax
  \def\bibfnamefont#1{#1}\fi
\expandafter\ifx\csname citenamefont\endcsname\relax
  \def\citenamefont#1{#1}\fi
\expandafter\ifx\csname url\endcsname\relax
  \def\url#1{\texttt{#1}}\fi
\expandafter\ifx\csname urlprefix\endcsname\relax\def\urlprefix{URL }\fi
\providecommand{\bibinfo}[2]{#2}
\providecommand{\eprint}[2][]{\url{#2}}

\bibitem[{\citenamefont{Reitze et~al.}(2019)\citenamefont{Reitze, Adhikari,
  Ballmer, Barish, Barsotti, Billingsley, Brown, Chen, Coyne, Eisenstein
  et~al.}}]{reitze2019cosmic}
\bibinfo{author}{\bibfnamefont{D.}~\bibnamefont{Reitze}},
  \bibinfo{author}{\bibfnamefont{R.~X.} \bibnamefont{Adhikari}},
  \bibinfo{author}{\bibfnamefont{S.}~\bibnamefont{Ballmer}},
  \bibinfo{author}{\bibfnamefont{B.}~\bibnamefont{Barish}},
  \bibinfo{author}{\bibfnamefont{L.}~\bibnamefont{Barsotti}},
  \bibinfo{author}{\bibfnamefont{G.}~\bibnamefont{Billingsley}},
  \bibinfo{author}{\bibfnamefont{D.~A.} \bibnamefont{Brown}},
  \bibinfo{author}{\bibfnamefont{Y.}~\bibnamefont{Chen}},
  \bibinfo{author}{\bibfnamefont{D.}~\bibnamefont{Coyne}},
  \bibinfo{author}{\bibfnamefont{R.}~\bibnamefont{Eisenstein}},
  \bibnamefont{et~al.}, \emph{\bibinfo{title}{Cosmic explorer: The u.s.
  contribution to gravitational-wave astronomy beyond ligo}}
  (\bibinfo{year}{2019}), \eprint{1907.04833}.

\bibitem[{\citenamefont{Maggiore et~al.}(2020)\citenamefont{Maggiore, Broeck,
  Bartolo, Belgacem, Bertacca, Bizouard, Branchesi, Clesse, Foffa,
  García-Bellido et~al.}}]{Maggiore2020}
\bibinfo{author}{\bibfnamefont{M.}~\bibnamefont{Maggiore}},
  \bibinfo{author}{\bibfnamefont{C.~V.~D.} \bibnamefont{Broeck}},
  \bibinfo{author}{\bibfnamefont{N.}~\bibnamefont{Bartolo}},
  \bibinfo{author}{\bibfnamefont{E.}~\bibnamefont{Belgacem}},
  \bibinfo{author}{\bibfnamefont{D.}~\bibnamefont{Bertacca}},
  \bibinfo{author}{\bibfnamefont{M.~A.} \bibnamefont{Bizouard}},
  \bibinfo{author}{\bibfnamefont{M.}~\bibnamefont{Branchesi}},
  \bibinfo{author}{\bibfnamefont{S.}~\bibnamefont{Clesse}},
  \bibinfo{author}{\bibfnamefont{S.}~\bibnamefont{Foffa}},
  \bibinfo{author}{\bibfnamefont{J.}~\bibnamefont{García-Bellido}},
  \bibnamefont{et~al.}, \bibinfo{journal}{Journal of Cosmology and
  Astroparticle Physics} \textbf{\bibinfo{volume}{2020}},
  \bibinfo{pages}{050–050} (\bibinfo{year}{2020}), ISSN
  \bibinfo{issn}{1475-7516},
  \urlprefix\url{http://dx.doi.org/10.1088/1475-7516/2020/03/050}.

\bibitem[{\citenamefont{Baibhav et~al.}(2019)\citenamefont{Baibhav, Berti,
  Gerosa, Mapelli, Giacobbo, Bouffanais, and Di~Carlo}}]{Baibhav_2019}
\bibinfo{author}{\bibfnamefont{V.}~\bibnamefont{Baibhav}},
  \bibinfo{author}{\bibfnamefont{E.}~\bibnamefont{Berti}},
  \bibinfo{author}{\bibfnamefont{D.}~\bibnamefont{Gerosa}},
  \bibinfo{author}{\bibfnamefont{M.}~\bibnamefont{Mapelli}},
  \bibinfo{author}{\bibfnamefont{N.}~\bibnamefont{Giacobbo}},
  \bibinfo{author}{\bibfnamefont{Y.}~\bibnamefont{Bouffanais}},
  \bibnamefont{and} \bibinfo{author}{\bibfnamefont{U.~N.}
  \bibnamefont{Di~Carlo}}, \bibinfo{journal}{Physical Review D}
  \textbf{\bibinfo{volume}{100}} (\bibinfo{year}{2019}), ISSN
  \bibinfo{issn}{2470-0029},
  \urlprefix\url{http://dx.doi.org/10.1103/PhysRevD.100.064060}.

\bibitem[{\citenamefont{Samajdar et~al.}(2021)\citenamefont{Samajdar, Janquart,
  van~den Broeck, and Dietrich}}]{Samajdar}
\bibinfo{author}{\bibfnamefont{A.}~\bibnamefont{Samajdar}},
  \bibinfo{author}{\bibfnamefont{J.}~\bibnamefont{Janquart}},
  \bibinfo{author}{\bibfnamefont{C.}~\bibnamefont{van~den Broeck}},
  \bibnamefont{and} \bibinfo{author}{\bibfnamefont{T.}~\bibnamefont{Dietrich}},
  \bibinfo{journal}{In Prep.}  (\bibinfo{year}{2021}),
  \bibinfo{note}{dCC/P2100030}.

\bibitem[{\citenamefont{Vitale and Evans}(2017)}]{Vitale_2017}
\bibinfo{author}{\bibfnamefont{S.}~\bibnamefont{Vitale}} \bibnamefont{and}
  \bibinfo{author}{\bibfnamefont{M.}~\bibnamefont{Evans}},
  \bibinfo{journal}{Physical Review D} \textbf{\bibinfo{volume}{95}}
  (\bibinfo{year}{2017}), ISSN \bibinfo{issn}{2470-0029},
  \urlprefix\url{http://dx.doi.org/10.1103/PhysRevD.95.064052}.

\bibitem[{\citenamefont{Adhikari et~al.}(2019)\citenamefont{Adhikari, Ajith,
  Chen, Clark, Dergachev, Fotopoulos, Gossan, Mandel, Okounkova, Raymond
  et~al.}}]{Adhikari_2019}
\bibinfo{author}{\bibfnamefont{R.~X.} \bibnamefont{Adhikari}},
  \bibinfo{author}{\bibfnamefont{P.}~\bibnamefont{Ajith}},
  \bibinfo{author}{\bibfnamefont{Y.}~\bibnamefont{Chen}},
  \bibinfo{author}{\bibfnamefont{J.~A.} \bibnamefont{Clark}},
  \bibinfo{author}{\bibfnamefont{V.}~\bibnamefont{Dergachev}},
  \bibinfo{author}{\bibfnamefont{N.~V.} \bibnamefont{Fotopoulos}},
  \bibinfo{author}{\bibfnamefont{S.~E.} \bibnamefont{Gossan}},
  \bibinfo{author}{\bibfnamefont{I.}~\bibnamefont{Mandel}},
  \bibinfo{author}{\bibfnamefont{M.}~\bibnamefont{Okounkova}},
  \bibinfo{author}{\bibfnamefont{V.}~\bibnamefont{Raymond}},
  \bibnamefont{et~al.}, \bibinfo{journal}{Classical and Quantum Gravity}
  \textbf{\bibinfo{volume}{36}}, \bibinfo{pages}{245010}
  (\bibinfo{year}{2019}), ISSN \bibinfo{issn}{1361-6382},
  \urlprefix\url{http://dx.doi.org/10.1088/1361-6382/ab3cff}.

\bibitem[{\citenamefont{Veitch et~al.}(2015)\citenamefont{Veitch, Raymond,
  Farr, Farr, Graff, Vitale, Aylott, Blackburn, Christensen, Coughlin
  et~al.}}]{Veitch_2015}
\bibinfo{author}{\bibfnamefont{J.}~\bibnamefont{Veitch}},
  \bibinfo{author}{\bibfnamefont{V.}~\bibnamefont{Raymond}},
  \bibinfo{author}{\bibfnamefont{B.}~\bibnamefont{Farr}},
  \bibinfo{author}{\bibfnamefont{W.}~\bibnamefont{Farr}},
  \bibinfo{author}{\bibfnamefont{P.}~\bibnamefont{Graff}},
  \bibinfo{author}{\bibfnamefont{S.}~\bibnamefont{Vitale}},
  \bibinfo{author}{\bibfnamefont{B.}~\bibnamefont{Aylott}},
  \bibinfo{author}{\bibfnamefont{K.}~\bibnamefont{Blackburn}},
  \bibinfo{author}{\bibfnamefont{N.}~\bibnamefont{Christensen}},
  \bibinfo{author}{\bibfnamefont{M.}~\bibnamefont{Coughlin}},
  \bibnamefont{et~al.}, \bibinfo{journal}{Physical Review D}
  \textbf{\bibinfo{volume}{91}} (\bibinfo{year}{2015}), ISSN
  \bibinfo{issn}{1550-2368},
  \urlprefix\url{http://dx.doi.org/10.1103/PhysRevD.91.042003}.

\bibitem[{\citenamefont{Ashton et~al.}(2019{\natexlab{a}})}]{Ashton:2018jfp}
\bibinfo{author}{\bibfnamefont{G.}~\bibnamefont{Ashton}} \bibnamefont{et~al.},
  \bibinfo{journal}{Astrophys. J. Suppl.} \textbf{\bibinfo{volume}{241}},
  \bibinfo{pages}{27} (\bibinfo{year}{2019}{\natexlab{a}}),
  \eprint{1811.02042}.

\bibitem[{\citenamefont{Vallisneri}(2008)}]{Vallisneri_2008}
\bibinfo{author}{\bibfnamefont{M.}~\bibnamefont{Vallisneri}},
  \bibinfo{journal}{Physical Review D} \textbf{\bibinfo{volume}{77}}
  (\bibinfo{year}{2008}), ISSN \bibinfo{issn}{1550-2368},
  \urlprefix\url{http://dx.doi.org/10.1103/PhysRevD.77.042001}.

\bibitem[{\citenamefont{Smith et~al.}(2020)\citenamefont{Smith, Ashton,
  Vajpeyi, and Talbot}}]{Smith_2020}
\bibinfo{author}{\bibfnamefont{R.~J.~E.} \bibnamefont{Smith}},
  \bibinfo{author}{\bibfnamefont{G.}~\bibnamefont{Ashton}},
  \bibinfo{author}{\bibfnamefont{A.}~\bibnamefont{Vajpeyi}}, \bibnamefont{and}
  \bibinfo{author}{\bibfnamefont{C.}~\bibnamefont{Talbot}},
  \bibinfo{journal}{Monthly Notices of the Royal Astronomical Society}
  \textbf{\bibinfo{volume}{498}}, \bibinfo{pages}{4492–4502}
  (\bibinfo{year}{2020}), ISSN \bibinfo{issn}{1365-2966},
  \urlprefix\url{http://dx.doi.org/10.1093/mnras/staa2483}.

\bibitem[{\citenamefont{Zhao and Wen}(2018)}]{Zhao_2018}
\bibinfo{author}{\bibfnamefont{W.}~\bibnamefont{Zhao}} \bibnamefont{and}
  \bibinfo{author}{\bibfnamefont{L.}~\bibnamefont{Wen}},
  \bibinfo{journal}{Physical Review D} \textbf{\bibinfo{volume}{97}}
  (\bibinfo{year}{2018}), ISSN \bibinfo{issn}{2470-0029},
  \urlprefix\url{http://dx.doi.org/10.1103/PhysRevD.97.064031}.

\bibitem[{\citenamefont{Canizares et~al.}(2015)\citenamefont{Canizares, Field,
  Gair, Raymond, Smith, and Tiglio}}]{Canizares:2014fya}
\bibinfo{author}{\bibfnamefont{P.}~\bibnamefont{Canizares}},
  \bibinfo{author}{\bibfnamefont{S.~E.} \bibnamefont{Field}},
  \bibinfo{author}{\bibfnamefont{J.}~\bibnamefont{Gair}},
  \bibinfo{author}{\bibfnamefont{V.}~\bibnamefont{Raymond}},
  \bibinfo{author}{\bibfnamefont{R.}~\bibnamefont{Smith}}, \bibnamefont{and}
  \bibinfo{author}{\bibfnamefont{M.}~\bibnamefont{Tiglio}},
  \bibinfo{journal}{Phys. Rev. Lett.} \textbf{\bibinfo{volume}{114}},
  \bibinfo{pages}{071104} (\bibinfo{year}{2015}), \eprint{1404.6284}.

\bibitem[{\citenamefont{Smith et~al.}(2016)\citenamefont{Smith, Field,
  Blackburn, Haster, P{\"{u}}rrer, Raymond, and Schmidt}}]{PhysRevD.94.044031}
\bibinfo{author}{\bibfnamefont{R.}~\bibnamefont{Smith}},
  \bibinfo{author}{\bibfnamefont{S.~E.} \bibnamefont{Field}},
  \bibinfo{author}{\bibfnamefont{K.}~\bibnamefont{Blackburn}},
  \bibinfo{author}{\bibfnamefont{C.-J.} \bibnamefont{Haster}},
  \bibinfo{author}{\bibfnamefont{M.}~\bibnamefont{P{\"{u}}rrer}},
  \bibinfo{author}{\bibfnamefont{V.}~\bibnamefont{Raymond}}, \bibnamefont{and}
  \bibinfo{author}{\bibfnamefont{P.}~\bibnamefont{Schmidt}},
  \bibinfo{journal}{Phys. Rev. D} \textbf{\bibinfo{volume}{94}},
  \bibinfo{pages}{44031} (\bibinfo{year}{2016}).

\bibitem[{\citenamefont{Pürrer}(2014)}]{Purrer:2014fza}
\bibinfo{author}{\bibfnamefont{M.}~\bibnamefont{Pürrer}},
  \bibinfo{journal}{Class. Quant. Grav.} \textbf{\bibinfo{volume}{31}},
  \bibinfo{pages}{195010} (\bibinfo{year}{2014}), \eprint{1402.4146}.

\bibitem[{\citenamefont{Lindblom et~al.}(2008)\citenamefont{Lindblom, Owen, and
  Brown}}]{PhysRevD.78.124020}
\bibinfo{author}{\bibfnamefont{L.}~\bibnamefont{Lindblom}},
  \bibinfo{author}{\bibfnamefont{B.~J.} \bibnamefont{Owen}}, \bibnamefont{and}
  \bibinfo{author}{\bibfnamefont{D.~A.} \bibnamefont{Brown}},
  \bibinfo{journal}{Phys. Rev. D} \textbf{\bibinfo{volume}{78}},
  \bibinfo{pages}{124020} (\bibinfo{year}{2008}),
  \urlprefix\url{https://link.aps.org/doi/10.1103/PhysRevD.78.124020}.

\bibitem[{\citenamefont{Antil et~al.}(2013)\citenamefont{Antil, Field,
  Herrmann, Nochetto, and Tiglio}}]{antil2013two}
\bibinfo{author}{\bibfnamefont{H.}~\bibnamefont{Antil}},
  \bibinfo{author}{\bibfnamefont{S.~E.} \bibnamefont{Field}},
  \bibinfo{author}{\bibfnamefont{F.}~\bibnamefont{Herrmann}},
  \bibinfo{author}{\bibfnamefont{R.~H.} \bibnamefont{Nochetto}},
  \bibnamefont{and} \bibinfo{author}{\bibfnamefont{M.}~\bibnamefont{Tiglio}},
  \bibinfo{journal}{Journal of Scientific Computing}
  \textbf{\bibinfo{volume}{57}}, \bibinfo{pages}{604} (\bibinfo{year}{2013}).

\bibitem[{\citenamefont{Abbott et~al.}(2017)}]{GW170817_stoch}
\bibinfo{author}{\bibfnamefont{B.~P.} \bibnamefont{Abbott}}
  \bibnamefont{et~al.}, \bibinfo{journal}{ArXiv e-prints}
  (\bibinfo{year}{2017}), \eprint{1710.05837}.

\bibitem[{\citenamefont{Owen and Sathyaprakash}(1999)}]{Owen1999}
\bibinfo{author}{\bibfnamefont{B.~J.} \bibnamefont{Owen}} \bibnamefont{and}
  \bibinfo{author}{\bibfnamefont{B.~S.} \bibnamefont{Sathyaprakash}},
  \bibinfo{journal}{Phys. Rev. D} \textbf{\bibinfo{volume}{60}},
  \bibinfo{pages}{022002} (\bibinfo{year}{1999}).

\bibitem[{\citenamefont{Abbott et~al.}(2020{\natexlab{a}})\citenamefont{Abbott,
  Abbott, Abbott, Abraham, Acernese, Ackley, Adams, Adya, Affeldt, and
  et~al.}}]{Abbott_2020_LRR}
\bibinfo{author}{\bibfnamefont{B.~P.} \bibnamefont{Abbott}},
  \bibinfo{author}{\bibfnamefont{R.}~\bibnamefont{Abbott}},
  \bibinfo{author}{\bibfnamefont{T.~D.} \bibnamefont{Abbott}},
  \bibinfo{author}{\bibfnamefont{S.}~\bibnamefont{Abraham}},
  \bibinfo{author}{\bibfnamefont{F.}~\bibnamefont{Acernese}},
  \bibinfo{author}{\bibfnamefont{K.}~\bibnamefont{Ackley}},
  \bibinfo{author}{\bibfnamefont{C.}~\bibnamefont{Adams}},
  \bibinfo{author}{\bibfnamefont{V.~B.} \bibnamefont{Adya}},
  \bibinfo{author}{\bibfnamefont{C.}~\bibnamefont{Affeldt}}, \bibnamefont{and}
  \bibinfo{author}{\bibnamefont{et~al.}}, \bibinfo{journal}{Living Reviews in
  Relativity} \textbf{\bibinfo{volume}{23}}
  (\bibinfo{year}{2020}{\natexlab{a}}), ISSN \bibinfo{issn}{1433-8351},
  \urlprefix\url{http://dx.doi.org/10.1007/s41114-020-00026-9}.

\bibitem[{\citenamefont{Veitch and Vecchio}(2010)}]{PhysRevD.81.062003}
\bibinfo{author}{\bibfnamefont{J.}~\bibnamefont{Veitch}} \bibnamefont{and}
  \bibinfo{author}{\bibfnamefont{A.}~\bibnamefont{Vecchio}},
  \bibinfo{journal}{Phys. Rev. D} \textbf{\bibinfo{volume}{81}},
  \bibinfo{pages}{062003} (\bibinfo{year}{2010}),
  \urlprefix\url{https://link.aps.org/doi/10.1103/PhysRevD.81.062003}.

\bibitem[{\citenamefont{Antil et~al.}(2018)\citenamefont{Antil, Chen, and
  Field}}]{greedycpp}
\bibinfo{author}{\bibfnamefont{H.}~\bibnamefont{Antil}},
  \bibinfo{author}{\bibfnamefont{D.}~\bibnamefont{Chen}}, \bibnamefont{and}
  \bibinfo{author}{\bibfnamefont{S.}~\bibnamefont{Field}},
  \bibinfo{journal}{Computing in Science $\&$ Engineering}
  \textbf{\bibinfo{volume}{20}}, \bibinfo{pages}{10–25}
  (\bibinfo{year}{2018}), ISSN \bibinfo{issn}{1558-366X},
  \urlprefix\url{http://dx.doi.org/10.1109/MCSE.2018.042781323}.

\bibitem[{\citenamefont{Field et~al.}(2014)\citenamefont{Field, Galley,
  Hesthaven, Kaye, and Tiglio}}]{PhysRevX.4.031006}
\bibinfo{author}{\bibfnamefont{S.~E.} \bibnamefont{Field}},
  \bibinfo{author}{\bibfnamefont{C.~R.} \bibnamefont{Galley}},
  \bibinfo{author}{\bibfnamefont{J.~S.} \bibnamefont{Hesthaven}},
  \bibinfo{author}{\bibfnamefont{J.}~\bibnamefont{Kaye}}, \bibnamefont{and}
  \bibinfo{author}{\bibfnamefont{M.}~\bibnamefont{Tiglio}},
  \bibinfo{journal}{Phys. Rev. X} \textbf{\bibinfo{volume}{4}},
  \bibinfo{pages}{031006} (\bibinfo{year}{2014}),
  \urlprefix\url{https://link.aps.org/doi/10.1103/PhysRevX.4.031006}.

\bibitem[{\citenamefont{Barrault et~al.}(2004)\citenamefont{Barrault, Maday,
  Nguyen, and Patera}}]{BARRAULT2004667}
\bibinfo{author}{\bibfnamefont{M.}~\bibnamefont{Barrault}},
  \bibinfo{author}{\bibfnamefont{Y.}~\bibnamefont{Maday}},
  \bibinfo{author}{\bibfnamefont{N.~C.} \bibnamefont{Nguyen}},
  \bibnamefont{and} \bibinfo{author}{\bibfnamefont{A.~T.}
  \bibnamefont{Patera}}, \bibinfo{journal}{Comptes Rendus Mathematique}
  \textbf{\bibinfo{volume}{339}}, \bibinfo{pages}{667 } (\bibinfo{year}{2004}),
  ISSN \bibinfo{issn}{1631-073X},
  \urlprefix\url{http://www.sciencedirect.com/science/article/pii/S1631073X04004248}.

\bibitem[{\citenamefont{Morisaki and Raymond}(2020)}]{Morisaki_2020}
\bibinfo{author}{\bibfnamefont{S.}~\bibnamefont{Morisaki}} \bibnamefont{and}
  \bibinfo{author}{\bibfnamefont{V.}~\bibnamefont{Raymond}},
  \bibinfo{journal}{Physical Review D} \textbf{\bibinfo{volume}{102}}
  (\bibinfo{year}{2020}), ISSN \bibinfo{issn}{2470-0029},
  \urlprefix\url{http://dx.doi.org/10.1103/PhysRevD.102.104020}.

\bibitem[{\citenamefont{Dietrich
  et~al.}(2019{\natexlab{a}})\citenamefont{Dietrich, Khan, Dudi, Kapadia,
  Kumar, Nagar, Ohme, Pannarale, Samajdar, Bernuzzi
  et~al.}}]{PhysRevD.99.024029}
\bibinfo{author}{\bibfnamefont{T.}~\bibnamefont{Dietrich}},
  \bibinfo{author}{\bibfnamefont{S.}~\bibnamefont{Khan}},
  \bibinfo{author}{\bibfnamefont{R.}~\bibnamefont{Dudi}},
  \bibinfo{author}{\bibfnamefont{S.~J.} \bibnamefont{Kapadia}},
  \bibinfo{author}{\bibfnamefont{P.}~\bibnamefont{Kumar}},
  \bibinfo{author}{\bibfnamefont{A.}~\bibnamefont{Nagar}},
  \bibinfo{author}{\bibfnamefont{F.}~\bibnamefont{Ohme}},
  \bibinfo{author}{\bibfnamefont{F.}~\bibnamefont{Pannarale}},
  \bibinfo{author}{\bibfnamefont{A.}~\bibnamefont{Samajdar}},
  \bibinfo{author}{\bibfnamefont{S.}~\bibnamefont{Bernuzzi}},
  \bibnamefont{et~al.}, \bibinfo{journal}{Phys. Rev. D}
  \textbf{\bibinfo{volume}{99}}, \bibinfo{pages}{024029}
  (\bibinfo{year}{2019}{\natexlab{a}}),
  \urlprefix\url{https://link.aps.org/doi/10.1103/PhysRevD.99.024029}.

\bibitem[{\citenamefont{Dietrich et~al.}(2017)\citenamefont{Dietrich, Bernuzzi,
  and Tichy}}]{PhysRevD.96.121501}
\bibinfo{author}{\bibfnamefont{T.}~\bibnamefont{Dietrich}},
  \bibinfo{author}{\bibfnamefont{S.}~\bibnamefont{Bernuzzi}}, \bibnamefont{and}
  \bibinfo{author}{\bibfnamefont{W.}~\bibnamefont{Tichy}},
  \bibinfo{journal}{Phys. Rev. D} \textbf{\bibinfo{volume}{96}},
  \bibinfo{pages}{121501} (\bibinfo{year}{2017}),
  \urlprefix\url{https://link.aps.org/doi/10.1103/PhysRevD.96.121501}.

\bibitem[{\citenamefont{Dietrich
  et~al.}(2019{\natexlab{b}})\citenamefont{Dietrich, Samajdar, Khan,
  Johnson-McDaniel, Dudi, and Tichy}}]{PhysRevD.100.044003}
\bibinfo{author}{\bibfnamefont{T.}~\bibnamefont{Dietrich}},
  \bibinfo{author}{\bibfnamefont{A.}~\bibnamefont{Samajdar}},
  \bibinfo{author}{\bibfnamefont{S.}~\bibnamefont{Khan}},
  \bibinfo{author}{\bibfnamefont{N.~K.} \bibnamefont{Johnson-McDaniel}},
  \bibinfo{author}{\bibfnamefont{R.}~\bibnamefont{Dudi}}, \bibnamefont{and}
  \bibinfo{author}{\bibfnamefont{W.}~\bibnamefont{Tichy}},
  \bibinfo{journal}{Phys. Rev. D} \textbf{\bibinfo{volume}{100}},
  \bibinfo{pages}{044003} (\bibinfo{year}{2019}{\natexlab{b}}),
  \urlprefix\url{https://link.aps.org/doi/10.1103/PhysRevD.100.044003}.

\bibitem[{\citenamefont{Borhanian}(2020)}]{Borhanian:2020ypi}
\bibinfo{author}{\bibfnamefont{S.}~\bibnamefont{Borhanian}}
  (\bibinfo{year}{2020}), \eprint{2010.15202}.

\bibitem[{\citenamefont{Poisson and Will}(1995)}]{Poisson_1995}
\bibinfo{author}{\bibfnamefont{E.}~\bibnamefont{Poisson}} \bibnamefont{and}
  \bibinfo{author}{\bibfnamefont{C.~M.} \bibnamefont{Will}},
  \bibinfo{journal}{Physical Review D} \textbf{\bibinfo{volume}{52}},
  \bibinfo{pages}{848–855} (\bibinfo{year}{1995}), ISSN
  \bibinfo{issn}{0556-2821},
  \urlprefix\url{http://dx.doi.org/10.1103/PhysRevD.52.848}.

\bibitem[{\citenamefont{Abbott et~al.}(2020{\natexlab{b}})\citenamefont{Abbott,
  Abbott, Abraham, Acernese, Ackley, Adams, Adhikari, Adya, Affeldt, Agathos
  et~al.}}]{190814}
\bibinfo{author}{\bibfnamefont{R.}~\bibnamefont{Abbott}},
  \bibinfo{author}{\bibfnamefont{T.~D.} \bibnamefont{Abbott}},
  \bibinfo{author}{\bibfnamefont{S.}~\bibnamefont{Abraham}},
  \bibinfo{author}{\bibfnamefont{F.}~\bibnamefont{Acernese}},
  \bibinfo{author}{\bibfnamefont{K.}~\bibnamefont{Ackley}},
  \bibinfo{author}{\bibfnamefont{C.}~\bibnamefont{Adams}},
  \bibinfo{author}{\bibfnamefont{R.~X.} \bibnamefont{Adhikari}},
  \bibinfo{author}{\bibfnamefont{V.~B.} \bibnamefont{Adya}},
  \bibinfo{author}{\bibfnamefont{C.}~\bibnamefont{Affeldt}},
  \bibinfo{author}{\bibfnamefont{M.}~\bibnamefont{Agathos}},
  \bibnamefont{et~al.}, \bibinfo{journal}{The Astrophysical Journal}
  \textbf{\bibinfo{volume}{896}}, \bibinfo{pages}{L44}
  (\bibinfo{year}{2020}{\natexlab{b}}), ISSN \bibinfo{issn}{2041-8213},
  \urlprefix\url{http://dx.doi.org/10.3847/2041-8213/ab960f}.

\bibitem[{\citenamefont{Essick et~al.}(2020)\citenamefont{Essick, Landry, and
  Holz}}]{Essick_2020}
\bibinfo{author}{\bibfnamefont{R.}~\bibnamefont{Essick}},
  \bibinfo{author}{\bibfnamefont{P.}~\bibnamefont{Landry}}, \bibnamefont{and}
  \bibinfo{author}{\bibfnamefont{D.~E.} \bibnamefont{Holz}},
  \bibinfo{journal}{Physical Review D} \textbf{\bibinfo{volume}{101}}
  (\bibinfo{year}{2020}), ISSN \bibinfo{issn}{2470-0029},
  \urlprefix\url{http://dx.doi.org/10.1103/PhysRevD.101.063007}.

\bibitem[{\citenamefont{{Lim} and {Holt}}(2019)}]{2019EPJA...55..209L}
\bibinfo{author}{\bibfnamefont{Y.}~\bibnamefont{{Lim}}} \bibnamefont{and}
  \bibinfo{author}{\bibfnamefont{J.~W.} \bibnamefont{{Holt}}},
  \bibinfo{journal}{European Physical Journal A} \textbf{\bibinfo{volume}{55}},
  \bibinfo{eid}{209} (\bibinfo{year}{2019}), \eprint{1902.05502}.

\bibitem[{\citenamefont{Ohme}(2012)}]{Ohme_2012}
\bibinfo{author}{\bibfnamefont{F.}~\bibnamefont{Ohme}},
  \bibinfo{journal}{Classical and Quantum Gravity}
  \textbf{\bibinfo{volume}{29}}, \bibinfo{pages}{124002}
  (\bibinfo{year}{2012}), ISSN \bibinfo{issn}{1361-6382},
  \urlprefix\url{http://dx.doi.org/10.1088/0264-9381/29/12/124002}.

\bibitem[{\citenamefont{Pankow et~al.}(2018)\citenamefont{Pankow,
  Chatziioannou, Chase, Littenberg, Evans, McIver, Cornish, Haster, Kanner,
  Raymond et~al.}}]{Pankow_2018}
\bibinfo{author}{\bibfnamefont{C.}~\bibnamefont{Pankow}},
  \bibinfo{author}{\bibfnamefont{K.}~\bibnamefont{Chatziioannou}},
  \bibinfo{author}{\bibfnamefont{E.~A.} \bibnamefont{Chase}},
  \bibinfo{author}{\bibfnamefont{T.~B.} \bibnamefont{Littenberg}},
  \bibinfo{author}{\bibfnamefont{M.}~\bibnamefont{Evans}},
  \bibinfo{author}{\bibfnamefont{J.}~\bibnamefont{McIver}},
  \bibinfo{author}{\bibfnamefont{N.~J.} \bibnamefont{Cornish}},
  \bibinfo{author}{\bibfnamefont{C.-J.} \bibnamefont{Haster}},
  \bibinfo{author}{\bibfnamefont{J.}~\bibnamefont{Kanner}},
  \bibinfo{author}{\bibfnamefont{V.}~\bibnamefont{Raymond}},
  \bibnamefont{et~al.}, \bibinfo{journal}{Physical Review D}
  \textbf{\bibinfo{volume}{98}} (\bibinfo{year}{2018}), ISSN
  \bibinfo{issn}{2470-0029},
  \urlprefix\url{http://dx.doi.org/10.1103/PhysRevD.98.084016}.

\bibitem[{\citenamefont{Ashton et~al.}(2019{\natexlab{b}})\citenamefont{Ashton,
  Hübner, Lasky, Talbot, Ackley, Biscoveanu, Chu, Divakarla, Easter, Goncharov
  et~al.}}]{bilby}
\bibinfo{author}{\bibfnamefont{G.}~\bibnamefont{Ashton}},
  \bibinfo{author}{\bibfnamefont{M.}~\bibnamefont{Hübner}},
  \bibinfo{author}{\bibfnamefont{P.~D.} \bibnamefont{Lasky}},
  \bibinfo{author}{\bibfnamefont{C.}~\bibnamefont{Talbot}},
  \bibinfo{author}{\bibfnamefont{K.}~\bibnamefont{Ackley}},
  \bibinfo{author}{\bibfnamefont{S.}~\bibnamefont{Biscoveanu}},
  \bibinfo{author}{\bibfnamefont{Q.}~\bibnamefont{Chu}},
  \bibinfo{author}{\bibfnamefont{A.}~\bibnamefont{Divakarla}},
  \bibinfo{author}{\bibfnamefont{P.~J.} \bibnamefont{Easter}},
  \bibinfo{author}{\bibfnamefont{B.}~\bibnamefont{Goncharov}},
  \bibnamefont{et~al.}, \bibinfo{journal}{The Astrophysical Journal Supplement
  Series} \textbf{\bibinfo{volume}{241}}, \bibinfo{pages}{27}
  (\bibinfo{year}{2019}{\natexlab{b}}), ISSN \bibinfo{issn}{1538-4365},
  \urlprefix\url{http://dx.doi.org/10.3847/1538-4365/ab06fc}.

\bibitem[{\citenamefont{Bustillo et~al.}(2020)\citenamefont{Bustillo, Dietrich,
  and Lasky}}]{bustillo2020higherorder}
\bibinfo{author}{\bibfnamefont{J.~C.} \bibnamefont{Bustillo}},
  \bibinfo{author}{\bibfnamefont{T.}~\bibnamefont{Dietrich}}, \bibnamefont{and}
  \bibinfo{author}{\bibfnamefont{P.~D.} \bibnamefont{Lasky}},
  \emph{\bibinfo{title}{Higher-order gravitational-wave modes will allow for
  percent-level measurements of hubble's constant with single binary neutron
  star merger observations}} (\bibinfo{year}{2020}), \eprint{2006.11525}.

\bibitem[{\citenamefont{Betancourt}(2018)}]{betancourt2018conceptual}
\bibinfo{author}{\bibfnamefont{M.}~\bibnamefont{Betancourt}},
  \emph{\bibinfo{title}{A conceptual introduction to hamiltonian monte carlo}}
  (\bibinfo{year}{2018}), \eprint{1701.02434}.

\bibitem[{\citenamefont{Gabbard et~al.}(2020)\citenamefont{Gabbard, Messenger,
  Heng, Tonolini, and Murray-Smith}}]{gabbard2020bayesian}
\bibinfo{author}{\bibfnamefont{H.}~\bibnamefont{Gabbard}},
  \bibinfo{author}{\bibfnamefont{C.}~\bibnamefont{Messenger}},
  \bibinfo{author}{\bibfnamefont{I.~S.} \bibnamefont{Heng}},
  \bibinfo{author}{\bibfnamefont{F.}~\bibnamefont{Tonolini}}, \bibnamefont{and}
  \bibinfo{author}{\bibfnamefont{R.}~\bibnamefont{Murray-Smith}},
  \emph{\bibinfo{title}{Bayesian parameter estimation using conditional
  variational autoencoders for gravitational-wave astronomy}}
  (\bibinfo{year}{2020}), \eprint{1909.06296}.

\bibitem[{\citenamefont{Shen et~al.}(2019)\citenamefont{Shen, Huerta, Zhao,
  Jennings, and Sharma}}]{shen2019deterministic}
\bibinfo{author}{\bibfnamefont{H.}~\bibnamefont{Shen}},
  \bibinfo{author}{\bibfnamefont{E.~A.} \bibnamefont{Huerta}},
  \bibinfo{author}{\bibfnamefont{Z.}~\bibnamefont{Zhao}},
  \bibinfo{author}{\bibfnamefont{E.}~\bibnamefont{Jennings}}, \bibnamefont{and}
  \bibinfo{author}{\bibfnamefont{H.}~\bibnamefont{Sharma}},
  \emph{\bibinfo{title}{Deterministic and bayesian neural networks for
  low-latency gravitational wave parameter estimation of binary black hole
  mergers}} (\bibinfo{year}{2019}), \eprint{1903.01998}.

\bibitem[{\citenamefont{Chua and Vallisneri}(2020)}]{PhysRevLett.124.041102}
\bibinfo{author}{\bibfnamefont{A.~J.~K.} \bibnamefont{Chua}} \bibnamefont{and}
  \bibinfo{author}{\bibfnamefont{M.}~\bibnamefont{Vallisneri}},
  \bibinfo{journal}{Phys. Rev. Lett.} \textbf{\bibinfo{volume}{124}},
  \bibinfo{pages}{041102} (\bibinfo{year}{2020}),
  \urlprefix\url{https://link.aps.org/doi/10.1103/PhysRevLett.124.041102}.

\bibitem[{\citenamefont{Chua et~al.}(2019)\citenamefont{Chua, Galley, and
  Vallisneri}}]{PhysRevLett.122.211101}
\bibinfo{author}{\bibfnamefont{A.~J.~K.} \bibnamefont{Chua}},
  \bibinfo{author}{\bibfnamefont{C.~R.} \bibnamefont{Galley}},
  \bibnamefont{and}
  \bibinfo{author}{\bibfnamefont{M.}~\bibnamefont{Vallisneri}},
  \bibinfo{journal}{Phys. Rev. Lett.} \textbf{\bibinfo{volume}{122}},
  \bibinfo{pages}{211101} (\bibinfo{year}{2019}),
  \urlprefix\url{https://link.aps.org/doi/10.1103/PhysRevLett.122.211101}.

\end{thebibliography}

\end{document}